\documentclass{article}
\usepackage[margin=0.5in]{geometry}
\usepackage{setspace}
\usepackage{amsmath}
\usepackage{mathptmx}      
\usepackage{hyperref}
\usepackage{adjustbox}
\usepackage{gensymb}
\usepackage{epstopdf}
\usepackage{graphicx}
\usepackage{hyperref}
\usepackage{tikz}
\usepackage{algorithm,amssymb}
\usepackage{algpseudocode}
\usepackage[english]{babel}
\usepackage{multirow}
\usepackage{blkarray}
\usepackage{pbox}
\usepackage{authblk}
\usepackage{gensymb}
\usepackage[justification=centering]{subfig}
\usepackage{float}

\usepackage{epstopdf}
\AppendGraphicsExtensions{.tif}

\usetikzlibrary{calc}

\usetikzlibrary{fit,positioning,arrows,automata,calc}
\tikzset{
	main/.style={circle, minimum size = 7mm, thick, draw =black!80, node distance = 10mm, align=center, text width=7mm},
	connect/.style={-latex, thick},
	box/.style={rectangle, draw=black!100, node distance = 10mm}
}

\begin{document}


\title{RapidHARe: A computationally inexpensive method for real-time human activity recognition from wearable sensors}

\author[1]{Roman Chereshnev}
\author[1]{Attila Kert\'esz-Farkas\thanks{Correspondence to akerteszfarkas@hse.ru}}
\affil[1]{Department of Data Analysis and Artificial Intelligence, Faculty of Computer Science, National Research University Higher School of Economics (HSE), Moscow, Russia}
\maketitle 
%
%
%

\begin{abstract}
Recent human activity recognition (HAR) methods, based on on-body inertial sensors, have achieved increasing performance; however, this is at the expense of longer CPU calculations and greater energy consumption. Therefore, these complex models might not be suitable for real-time prediction in mobile  systems, e.g., in elder-care support and long-term health-monitoring systems.  Here, we present a new method called RapidHARe for real-time human activity recognition based on modeling the distribution of a raw data in a half-second context window using dynamic Bayesian networks. Our method does not employ any dynamic-programming-based algorithms, which are notoriously slow for inference, nor does it employ feature extraction or selection methods. In our comparative tests, we show that RapidHARe is an extremely fast predictor, one and a half times faster than artificial neural networks (ANNs) methods, and more than eight times faster than recurrent neural networks (RNNs) and hidden Markov models (HMMs). Moreover, in performance, RapidHare achieves an F1 score of 94.27\% and accuracy of 98.94\%, and when compared to ANN, RNN, HMM, it reduces the F1-score error rate by 45\%, 65\%, and 63\% and the accuracy error rate by 41\%, 55\%, and 62\%, respectively. Therefore, RapidHARe is suitable for real-time recognition in mobile devices.

\end{abstract}

%
%

\section{Introduction}
\label{intro}
The increasing availability of wearable body sensors leads to novel scientific studies and industrial applications in the ubiquitous computing field \cite{aggarwal2013managing,shoaib2015survey}. The main areas include gesture recognition (GR) \cite{amma2014airwriting}, recognition of activities of daily living (ADL), human activity recognition (HAR) \cite{bao2004activity}, and human gait analysis (HGA) \cite{sant2011new}. Gesture recognition mainly focuses on recognizing hand-drawn gestures in the air. Patterns to be recognized may include numbers, circles, boxes, or Latin alphabet letters. Recognition of activities of daily living, on the other hand, aims to recognize daily lifestyle activities performed primarily by the subject's \cite{Chavarriaga2013,sagha2011benchmarking}. For instance, an interesting research topic is recognizing activities in or around the kitchen, such as cooking, loading the dishwasher or washing machine, and so on \cite{pham2009slice}. Often, these activities can be interrupted by, for example, answering the phone. Human activity recognition (HAR) usually focuses on activities related to or performed by legs, such as walking, jogging, turning left or right, jumping, lying down, going up or down the stairs, sitting down, and so on. Human gait analysis (HGA) focuses not only on the recognition of activities observed but also on how activities are performed. This can be useful in health-care systems for monitoring patients recovering after surgery, fall detection, or diagnosing the state of, for example, Parkinson's disease \cite{sant2011new,sant2012symbolic,Comber201725}. An important application in HGA is installing body accelerometers on the hips and legs of people with Parkinson's disease \cite{bachlin2010wearable}. Here, the objective is to detect freezing of the gait and prevent falling incidents. 

Our research group generally focuses on developing methods related to HGA and HAR, and in this article we were interested in and studied HAR methods, which have  the following properties:
\begin{enumerate}
	\itemsep0cm 
	\item Low prediction latency.
	\item Smooth, continuous activity recognition within a given activity and rapid transition in between different activities.
	\item Speed and energy efficiency for mobile-pervasive technologies.
\end{enumerate}

The first requirement ensures that the model is of low latency; therefore, activity prediction can be made instantly based on the latest observed data. Therefore, bidirectional models, such as bidirectional long short-term memory (LSTM) recurrent neural networks (RNN) \cite{lefebvre2015inertial} or dynamic time warping (DTW) \cite{liu2009uwave} methods, are not appropriate for our aims for two main reasons: First, these bidirectional methods require a whole observed sequence before making any predictions, which would therefore increase their latency. Second, the prediction they make on a frame is based on subsequent data. Standard hidden Markov models (HMMs) have become the {\em de facto} approach for activity recognition \cite{olguin2006human,mannini2010machine,lester2005hybrid,junker2008gesture}, and they yield good performance in general. However, they do so at the expense of increased latency in prediction, because Viterbi algorithms use the whole sequence, or at least some part of it, to estimate a series of activities (i.e., hidden states), and their time complexity is polynomial. Therefore, in our opinion, HMMs are not adequate for on-the-fly prediction, because the latency of these methods can be considered rather high.

The second point is to ensure that an activity recognition method provides consistent prediction within the same activity, but changes rapidly when the activity has changed. Lester et al. \cite{lester2005hybrid} have pointed out that a single-frame prediction method such as decision stumps is prone to yielding scattered predictions. However, human activity data are time series data in nature, and subsequent data frames are highly correlated. This tremendous amount of information can be exploited simply by sequential models such as HMM and RNN, or by incorporating the sliding-window technique to single-frame methods (e.g., nearest-neighbor). In fact, the authors in \cite{mannini2010machine} have pointed out that the continuous-emissions HMM-based sequential classifier (cHMM) performs systematically better than its simple single-frame Gaussian mixture model (GMM) counterpart (99.1\% vs. 92.2\% in accuracy). Actually, the proposed sequential classifier wins over all its tested single-frame competitors (the best single-frame classifier is the nearest mean (NM) classifier which achieves up to 98.5\% in accuracy). This highlights the relevance of exploiting the statistical correlation from human dynamics. 

Continuous sensing and evaluating CPU-intensive prediction methods rapidly deplete a mobile system's energy. Therefore, the third point requires a system to be energy-efficient enough for mobile-pervasive technologies. Several approaches have been introduced for this problem. Some methods aim to keep the number of necessary sensors low by adaptive selection \cite{zappi2008activity} or based on the activity performed \cite{gordon2012energy,yan2012energy,krause2005trading},  for accurate activity prediction. Other approaches aim to reduce the computational cost by feature selection \cite{anguita2012human}, feature learning \cite{plotz2011feature}, or  proposing computationally inexpensive prediction models such as C4.5, random forest \cite{mazilu2012online}, or decision trees \cite{skotte2014detection}.  In this study, we put emphasis on a computationally inexpensive prediction model that uses little memory and takes few computational steps while still achieving good performance.

Recently, deep-learning technologies, deep LSTM, and deep convolutional LSTM (DCSLTM) have emerged for activity recognition systems with superb performance, mainly in ADL and GR \cite{ordonez2016deep,hammerla2016deep}. These methods are capable of learning features automatically from the data \cite{bengio2009learning}. The price of this skill is that they consist of millions of model parameters that are more difficult to train, and most importantly, they result in longer prediction times and require more CPU time compared to inexpensive models such as decision trees.  On the other hand, we argue that these methods have too high of a capacity for HAR and HGA problems, and thus they overfit. In our opinion, these problems involve only a few thousands input features, and the ``complexity" of the underlying data manifold is rather low.  LSTM methods have the capacity to remember the activity performed sometime ago, which might be useful for recognizing daily activities, such as scrambling eggs or washing dishes. However, for HAR and HGA-related problems, such skills are not needed, because we think that the current activity is independent of activities performed some time ago. For instance, if the next activity is going to be walking up the stairs, then it is because there are stairs ahead, and this fact is independent of previous activities, whether the user was sitting or running before. In our opinion, our hypothesis is supported by the studies in \cite{ordonez2016deep,hammerla2016deep}. Both studies have reported improvement in performance for ADL using deep LSTM methods. However, in freezing-of-gait prediction tasks, Hammerla et al. have reported a 76\% F1 score in Table 2 in \cite{hammerla2016deep}, while a simple method such as random forests and C4.5 using smartly crafted features has achieved an F1 score over 95\% on the same dataset, as shown in Table 2 in \cite{mazilu2012online}. Similar conclusions can be reached from the results presented in Table 2 in \cite{weiss2012impact}, where the nearest-neighbor and random forest methods outperform multilayer perceptrons in test scenarios (which the authors termed ``impersonal'' and ``hybrid'') in which training and test data were recorded by different users.  We think these results support our argument, and,  therefore, deep models of high capacity for HAR and HGA problems do not seem to be justified to us. We believe that smartly designed features used along with computationally inexpensive models can provide faster and more energy-efficient methods with low latency for this field.

In this article, we present a novel method for HAR called {\em RapidHARe} for real-time prediction of continuous activity recognition. The proposed model is a small dynamic Bayesian network that does not utilize the Viterbi algorithm or other dynamic programming approaches for activity prediction, but instead utilizes the data distribution within a small, half-second-long context window. Moreover, our method does not employ feature transformation and selection methods. This provides a quick method that does not require exhaustive CPU calculations.  Therefore, RapidHARe is suitable for real-time recognition. Moreover, it is inexpensive for mobile systems and can be employed in elder-care support and long-term health-monitoring systems such as freeze-of-gait prediction, fall detection, robotic exoskeletons in health care, and surgery recovery. 

This article is organized as follows: In section 2, we introduce the mathematical model of RapidHARe by using dynamic Bayesian networks. In section 3, we describe the data we used in our experiments. In section \ref{ch:results}, we present our experimental results obtained and discuss our findings. Finally, we conclude our study in the last section.

\section{Methods}
\label{metods}
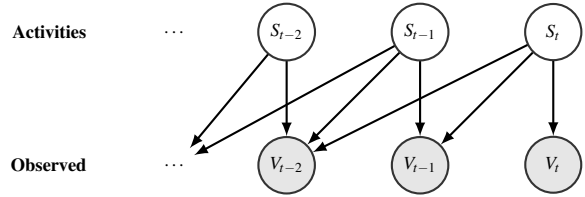
\begin{figure}[H]
	\centering
	\begin{tikzpicture}[thick,scale=0.7, every node/.style={scale=0.7}]
	\node[box,draw=white!100] (Latent) {\textbf{Activities}};
	\node[box,draw=white!100] (Space1) [right=of Latent] {\textbf{          }};
	\node[main] (L1) [right=of Space1] {$S_{t-2}$};
	\node[main] (L2) [right=of L1] {$S_{t-1}$};
	\node[main] (L3) [right=of L2] {$S_{t}$};
	\node[main,fill=black!10] (O1) [below=of L1] {$V_{t-2}$};
	\node[main,fill=black!10] (O2) [below=of L2] {$V_{t-1}$};
	\node[main,fill=black!10] (O3) [below=of L3] {$V_t$};
	\node[box,draw=white!100,left=of O1] (Space2) {\textbf{         }};
	\node[box,draw=white!100,left=of Space2] (Observed) {\textbf{Observed}};

	\path (Latent)  -- node[auto=false]{\dots} (L1);
	\path (Observed)  -- node[auto=false]{\dots} (O1);
	
	\path[->,shorten >= 5pt] (L1) edge [connect] (Space2);
	\path (L1) edge [connect] (O1);
	\path (L2) edge [connect] (O1);
	\path (L2) edge [connect] (O2);
	\path[->,shorten >= 5pt] (L2) edge [connect] (Space2);
	\path (L3) edge [connect] (O3);
	\path (L3) edge [connect] (O2);
	\path (L3) edge [connect] (O1);
	\end{tikzpicture}
	\caption{Illustration of an unfolded dynamic Bayesian network  w.r.t. an activity series.}
	\label{fg:Bayesian-Net}
\end{figure}

We created a dynamic Bayesian network, whose structure is shown in Figure \ref{fg:Bayesian-Net}. The states, i.e., activities, denoted by $S$ and the probability of a state $s_t$ at a given time $t$ with respect to a given observed context window $v_t, v_{t-1}, \cdots, v_{t-K}$ of length $K$, is formulated by 
\begin{equation}
\begin{gathered}
P(s_t\mid v_t, v_{t-1},  \cdots, v_{t-K}) = \\
{\prod_{k=0}^{K} P(v_{t-k} \mid s_t) P(s_t) \over 
	\sum_{n=1}^N \prod_{k=0}^{K} P(v_{t-k} \mid s_t = n)P(s_t=n)}.
\label{eq:stateprob}
\end{gathered}
\end{equation} 
Certainly, at the beginning of performance, when $t<K$, the context window is adjusted.
In our experiments, we did not use different {\em a priori} class probabilities for different $P(s_k)$. This is because we did not want our model to be biased toward some states that are abundant in the training data. Therefore, the activity prediction should be based fully on the data, and the state probabilities $P(s_k)$ can be omitted from Eq. \ref{eq:stateprob}.

The state being performed at time $t$ can be predicted as follows:
\begin{equation}
\hat{s}_t = \text{argmax}_{s_t} \left\{P(s_t\mid v_t, v_{t-1},  \cdots, v_{t-K})\right\}.
\label{eq:argmax1}
\end{equation}

Since the optimum of Eq. \ref{eq:argmax1} is invariant to normalization, the normalization factor can be omitted from Eq. \ref{eq:stateprob}. This gives us a very simple model for activity prediction in the following form:
\begin{equation}
\hat{s}_t = \text{argmax}_{s_t} \left\{\prod_{k=0}^{K} P(v_{t-k} \mid s_t)\right\}.
\label{eq:argmax2}
\end{equation}
This model can be implemented using the rolling-window technique for real-time continuous activity recognition; thus, the model remains fast for large $K$s, and redundant calculation of $P(v_{t-k}\mid s_t)\;(k>0)$ can be avoided by using tables. 

The distribution $P(V\mid S)$ with respect to a given state is modeled with Gaussian mixture models (GMMs), and its parameters are trained using the expectation-maximization (EM) method. The training of GMMs was straightforward because training data were segmented. 

Overall, we obtained a simple and fast model that consumes little energy to recognize human activities.

\begin{table*}[btp]
	\begin{center}
		\caption{Characteristics of data and activities}
		\label{tb:activity-data-list}
		\begin{tabular}{lllll}
			\hline
			Activity &Time sec (min) &Percent &Samples &Description\\
			\hline
			Walking    		& 5604 (93) 	& 27.66 & 314775 & \parbox{6cm}{ Walking and turning at various speeds on a flat surface } \\
			Running    		& 1141 (19) 	& 5.63 	& 64122  	 & Running at various paces\\
			Going up   		& 2343 (39) 	& 11.56 & 131604  & Going up stairs at various speeds\\
			Going down 		& 2076 (34) 	& 10.25 & 116637   & \parbox{6cm}{Going down stairs at various speeds}\\
			Sitting    		& 1336 (22) 	& 6.59 	& 75036    & Sitting on chair; floor not included\\
			Sitting down 	& 429  (7) 		& 2.12 	& 24112  	 & Sitting down on chair; floor not included\\
			Standing up 	& 398  (6) 		& 1.97 	& 22373	 & Standing up from a chair\\
			Standing   		& 6933 (115) 	& 34.22 & 389420 & Static standing on a solid surface\\
			\hline
			Total &20260 (335) &100.0 &11380793 \\
			\hline
		\end{tabular}
	\end{center}
\end{table*}

\section{Data collection}
\label{ch:data_acqusition}

\begin{figure}[tbp] 
	\centering
	\includegraphics[width=6cm,height=6cm]{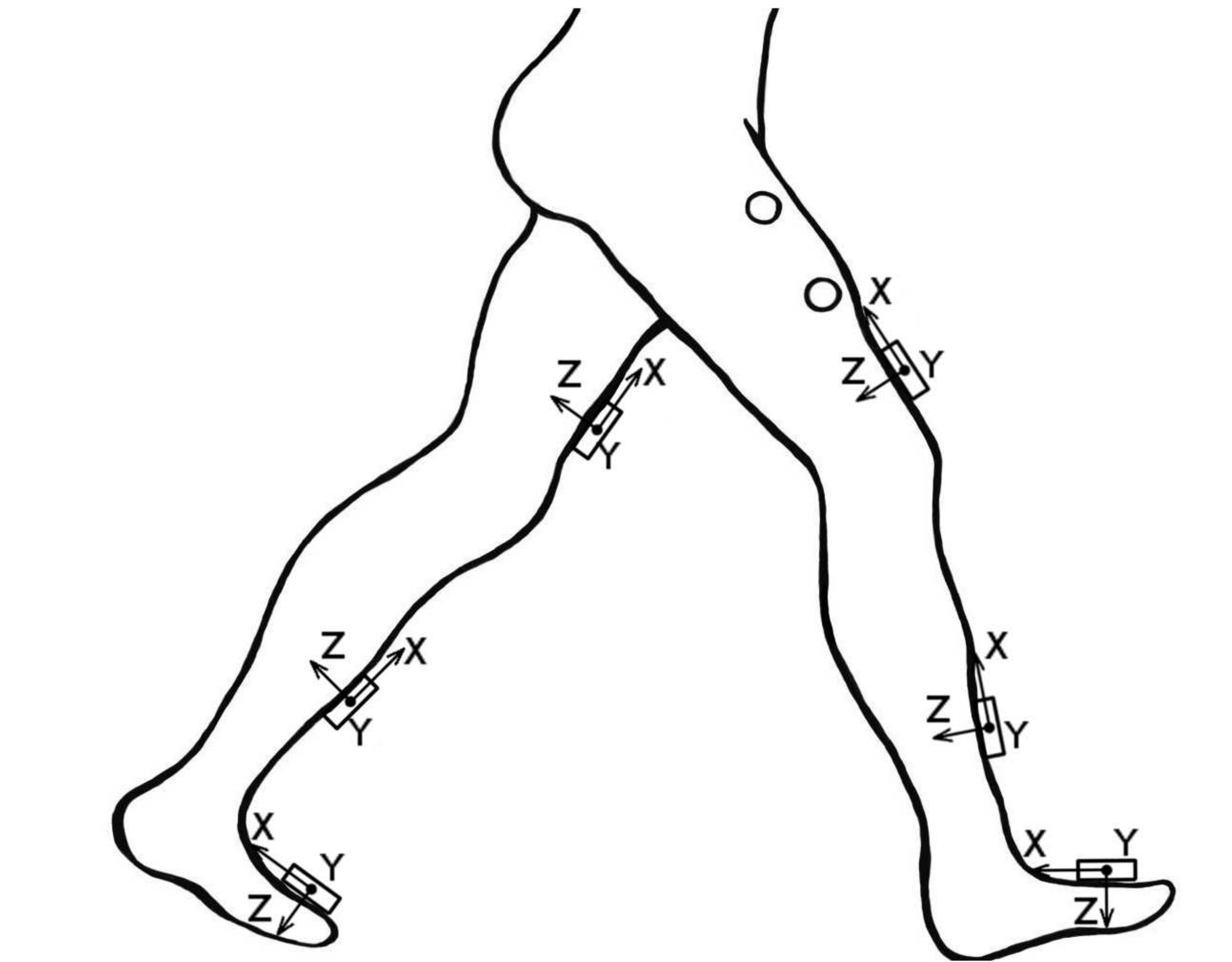}
	\caption{Sensor locations. Circles show EMG sensors, while boxes represent accelerometers and gyroscopes.}
	\label{fg:walking-scheme}
\end{figure}

To perform our experiments, we have recorded a total of 5 hours of data from 18 participants performing 8 different activities. These participants were healthy young adults: 4 females and 14 males with an average age of 23.67 years (standard deviation [STD]: 3.69), an average height of 179.06 cm (STD: 9.85), and an average weight of 73.44 kg (STD: 16.67). The participants performed a combination of activities at normal speed in a casual way, and there were no obstacles placed in their way.  For instance, starting in the sitting position, the participant was instructed to perform the following activities: sitting, standing up, walking, going up the stairs, walking, sitting down. The experimenter recorded the data continually using a laptop and annotated the data with the activities performed. This provided us a long, continuous sequence of segmented data annotated with activities. We developed our own data-collector program. In total, 1,138,079 samples were collected. A summary of the activities recorded and other characteristics of the data is shown in Table \ref{tb:activity-data-list}.

During data collection, we used MPU9250 inertial sensors and electromyography (EMG) sensors made in the Laboratory of Applied Cybernetics Systems, MIPT (\url{www.mipt.ru}).
Each EMG sensor has a voltage gain of about 5000, and a band-pass filter with bandwidth corresponding to a power spectrum of EMG (10--500 Hz). The sample rate of each EMG-channel is 1.0 kHz, the ADC resolution is 8 bits, and the input voltages is 0--5 V.
The inertial sensors consisted of a three-axis accelerometer and a three-axis gyroscope integrated into a single chip. Data were collected with the accelerometer's range equal to $\pm 2g$ with sensitivity 16.384 LSB/g and the gyroscope's range equal to $\pm 2000 \degree/$s with sensitivity 16.4 LSB $/ \degree/$s. 
All sensors were powered with a battery, which helped to minimize electrical grid noise. 

Accelerometer and gyroscope signals were stored in int16 format. EMG signals were stored in uint8. In our experiments, all data were scaled to range $[-1,1]$. 

In total, six pieces of inertial sensors (three-axis accelerometer and three-axis gyroscope) and one pair of EMG sensors were installed symmetrically on the right and left legs with elastic bands. A pair of inertial sensors were installed on the rectus femoris muscle 5 cm above the knee, a pair of sensors around the middle of the shinbone at the level where the calf muscle ends, and a pair on the feet on the metatarsal bones. This provided 36 features. Two EMG sensors were placed on the vastus lateralis and connected to the skin by three electrodes. The EMG sensors additionally provided two more features.  The locations of the sensors are shown in Figure \ref{fg:walking-scheme}. In total, 38 signals were collected.

The sensors were connected through wires with each other and to a microcontroller box, which contained an Arduino electronics platform with a Bluetooth module. The microcontroller collected 56.3500 samples per second on average, with a STD 3.2057, and then transmitted them to a laptop through the Bluetooth connection.

Data acquisition was carried out mainly inside a building. We note that data were not recorded on a treadmill. The data are available in \cite{chereshnev2017HuGADB}.

\section{Results and discussions}
\label{ch:results}
The performance of our RapidHARe model was evaluated using a supervised cross-validation approach \cite{kertesz2008benchmarking}. In this approach, data from a designated  participant were held out for tests, data from another participant were held for validation, and the rest of the data from the 16 participants were used for training. Thus, this approach gives a reliable estimation of how an activity recognition system would perform on a new user whose data have not been seen before. In our experiments, we repeated this test for every user in the dataset and averaged the results. A similar testing procedure has been introduced by Weiss et al. \cite{weiss2012impact}. Our methods were implemented using the Python scikit-learn package  (version 0.18.1) on a PC equipped with Intel Core i7-4790 CPU, 8 Gb DDR-III 2400 MHz RAM, and Nvidia GTX Titan X GPU.

Please note that, besides the feature scaling described in section \ref{ch:data_acqusition}, we did not use any preprocessing step, feature extraction, or feature selection methods.  

\subsection{On hyperparameters}

\begin{figure}
		\centering
	\includegraphics[trim=0.2cm 0.0cm 0.0cm 0.0cm,clip=true, width=0.60\textwidth, height=8cm]{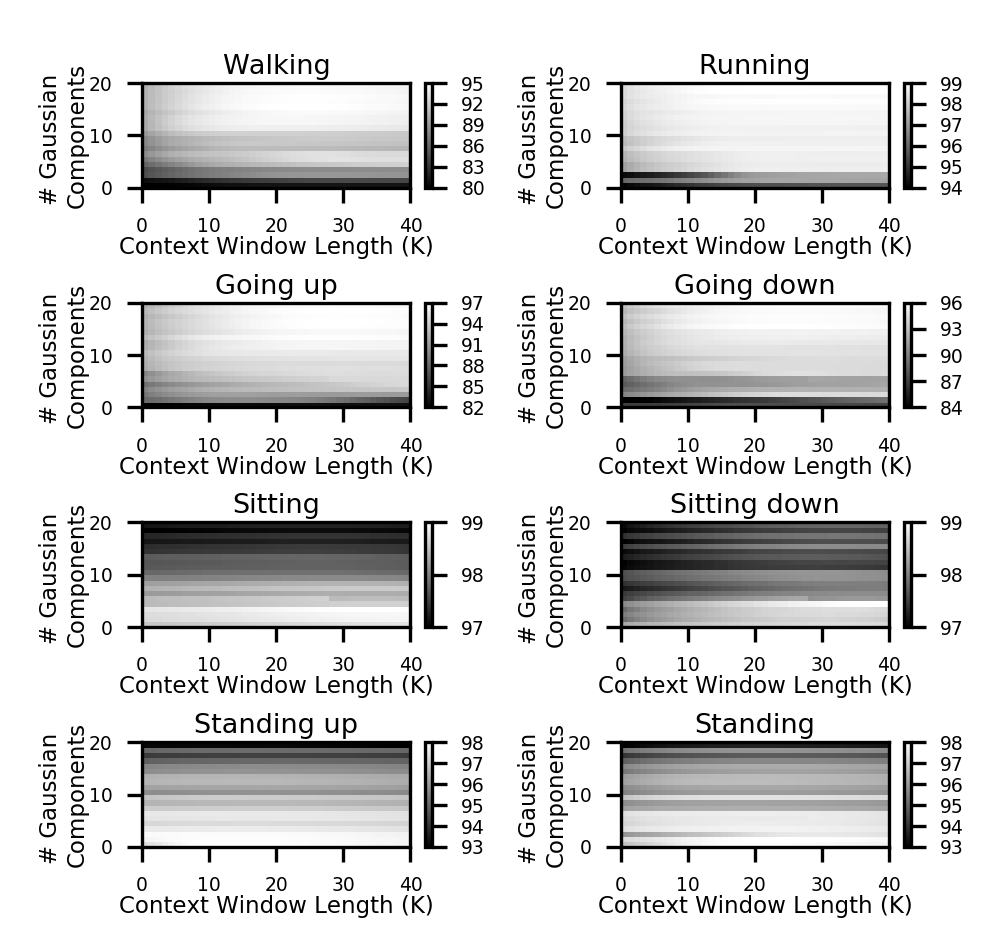} 
	\caption{Accuracy w.r.t. the number of Gaussian components and the length of the context window.}
	
	\label{fg:GMMComponentsA}
\end{figure}

\begin{figure}
		\centering
	\includegraphics[trim=0.2cm 0.0cm 0.0cm 0.0cm,clip=true, width=0.60\textwidth, height=8cm]{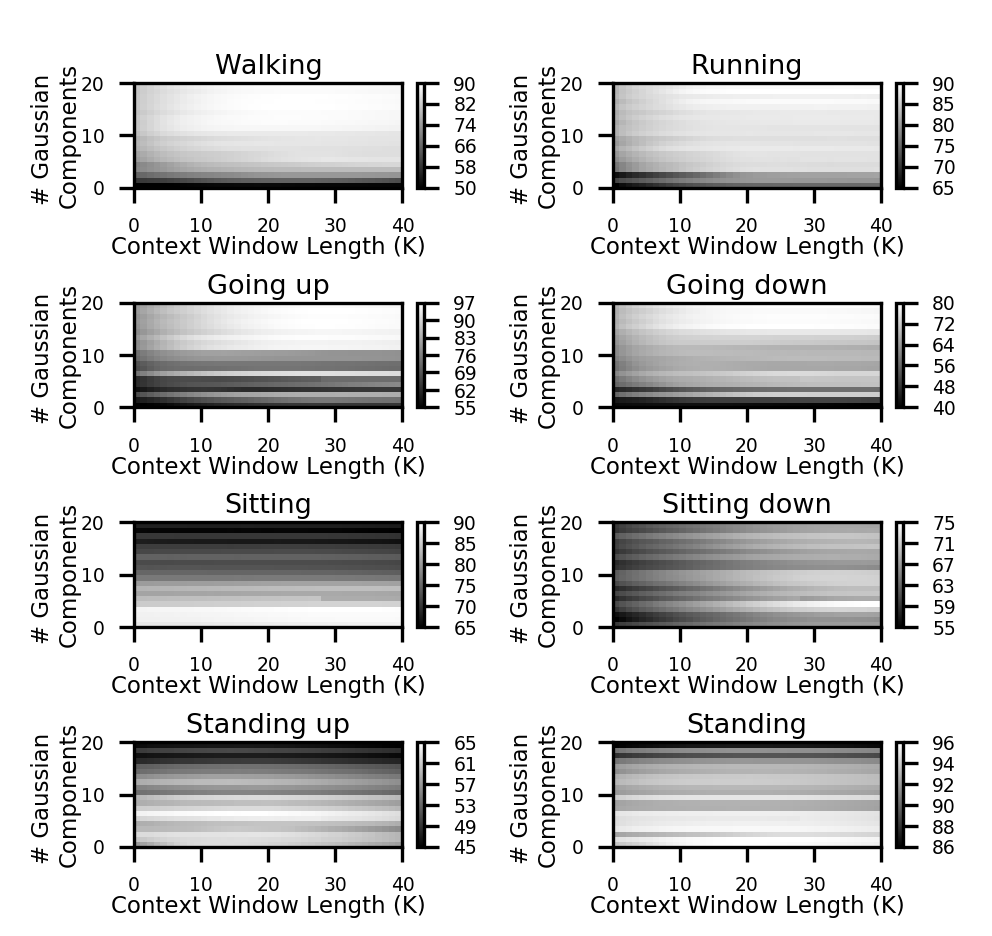} 
	\caption{F1 scores w.r.t. the number of Gaussian components and the length of the context window.}
	\label{fg:GMMComponentsF1}
\end{figure}

In our first experiment, we determined the values of the length of the context window and number of the Gaussian components in $P(V\mid S)$ via grid search for RapidHARe. In our tests, the covariance matrices $\Sigma$ in all Gaussian components were restricted to be diagonal. The results were evaluated in terms of accuracy and F1 score and are shown in Figures \ref{fg:GMMComponentsA} and \ref{fg:GMMComponentsF1}. They indicate that a good performance can be achieved using $K=26$ for the context window length. However, for the Gaussian components, it seems that for dynamic activities, such as walking and running, the higher the number of Gaussian components, the better the performance. On the other hand, for static activities, such as sitting and standing, a large number of Gaussian components hinders the activity recognition. Therefore, we set the number of Gaussian components for $P(V\mid S)$ for the following activities: walking, 18; running, 18; going up, 16; going down, 16; sitting, 2; standing up, 5; sitting down, 7; and standing, 4. The activity recognition results using these hyperparameters are shown in Table \ref{tb:hyperparam_res}, and we achieved 97.85\% accuracy, 87.4\% precision, 87.22\% recall, and an 86.4\% F1 score. The confusion matrix is shown in Table \ref{tb:hyperparam_conf}.

\begin{table*}[h]
	\centering
	\caption{Results of activity recognition}
	\label{tb:hyperparam_res}
	\begin{adjustbox}{max width=\textwidth}
		\begin{tabular}{c|ccccccccc}
			~& Walking & Running & Going up & Going down & Sitting & Sitting down & Standing up & Standing & Average\\ 
			\hline
			Recall     & 89.09  & 96.05  & 92.31   &  89.14    & 97.61  &   68.62     &   72.51    & 93.75   &       87.22  \\
			Precision  & 94.57  & 91.52  & 81.23   &  86.25    & 88.83  &   86.57     &   73.25    & 97.26   &        87.4    \\
			F1 score   & 91.62  & 93.18  & 85.8    &  87.25    & 92.89  &   75.05     &   70.95    & 95.35   &        86.4   \\
			Accuracy   & 95.63  & 99.32  & 96.37   &  97.24    & 98.97  &   99.03     &   98.76    & 97.66   &       97.85  \\			
		\end{tabular}
	\end{adjustbox}
\end{table*}

\begin{table*}[h]
	\centering
	\caption{Confusion matrix}
	\label{tb:hyperparam_conf}
	\begin{adjustbox}{max width=\textwidth}
		\begin{tabular}{c|cccccccc}			
			& Walking & Running & Going up & Going down & Sitting & Sitting down & Standing up & Standing \\
			\hline
			Walking   		& 854097 &  6702  & 43981  &  36469   &   0    &     34     &     89    &   2953  \\
			Running   		&  2401  & 186493 &  1533  &   1764   &   0    &     0      &     0     &   175   \\
			Going up 		& 18209  &  1677  & 363541 &  10560   &   0    &     61     &     1     &   763   \\
			Going down  	& 14190  &  6590  & 13266  &  314762  &   0    &     0      &     2     &   1101  \\
			Sitting  		&   0    &   0    &   0    &   586    & 218933 &    2244    &    3345   &    0    \\
			Sitting down 	&   0    &   0    &   0    &    0     & 11400  &   48866    &    2401   &   9669  \\
			Standing up  	&   0    &   0    &   0    &    2     & 14257  &    1033    &   48691   &   3136  \\
			Standing  		& 11149  &   72   & 23231  &   4704   &   0    &    5328    &   14198   & 1109578 \\		
		\end{tabular}
	\end{adjustbox}			
\end{table*}

\subsection{Continuous activity recognition}

\begin{figure*}[h]
	\includegraphics[width=\textwidth, height=25mm]{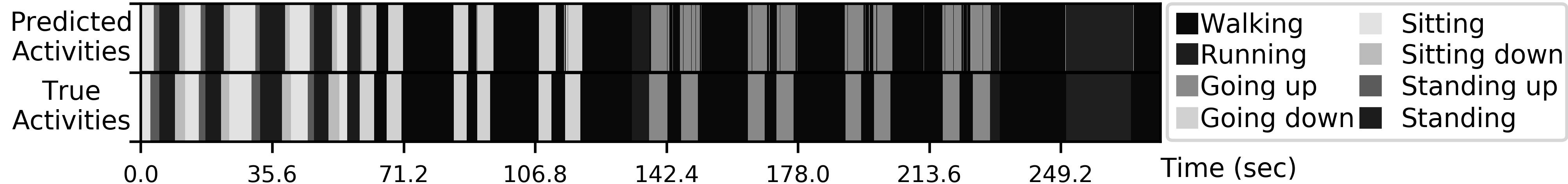}\\
	\caption{Continuous activity recognition}
	\label{fg:Activity_timelineA}
\end{figure*}

Next, we examined how well RapidHARe performs on continuous activity recognition. For this reason, we took a continuous series of activities and performed the activity recognition. Then, we plotted the true and predicted activities on a time line, shown in Figure \ref{fg:Activity_timelineA}.  The results show that our method does predict continuous activities, and it does not predict scattered activities for neighboring frames except for a few frames. 

\begin{figure*}[h]
	\includegraphics[width=\textwidth, height=25mm]{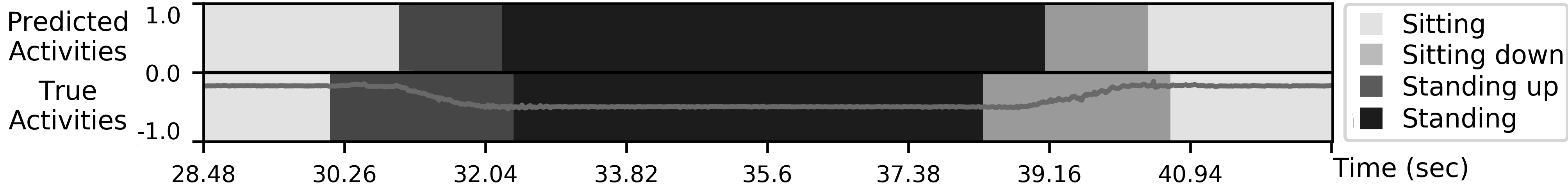}\\	
	\caption{Activity recognition at 35.6 s enlarged from Figure \ref{fg:Activity_timelineA}. The line represents the x-axis acceleration value recorded by accelerometer located on thigh.}
	\label{fg:Activity_timelineB}
\end{figure*}

However, it looks like, misclassification occurs on the borders in many cases. Furthermore, if we enlarge the standing--sitting activity at 35.6 sec, as shown in Figure \ref{fg:Activity_timelineB}, we can see that our method predicts sitting activity, at around 40.94 sec, a small fraction of a second earlier than it happened, according to the data annotation. It is unlikely that our method can predict the future. This phenomenon could be a result of inaccurate data segmentation made by the data controller and by the fact that it is difficult to exactly determine an activity border in 10--20 ms. We also plotted over the activities the signals measured by the x-axis accelerometer placed on the right thigh. This example shows that, in our opinion, the activity borders predicted by our model are actually aligned with the signal changes more appropriately than are the borders determined by the experimenter.

In order to mitigate this phenomenon, we allow some tolerance in the misclassification if it occurs on the activity border. Thus, we tolerate up to 25 data frames (which is about half a second) to be misclassified on the activity border if and only if our method correctly recognizes the succeeding activity. 
We believe that a half-second misclassification on the activity borders during continuous activity recognition is acceptable in practice. Moreover, if we allow misclassification on the borders, then we think the performance measures will put an emphasis on more reliable estimation for the actual scattered  misclassification made by the model, and it will be more tolerant of inaccurate data segmentation. 

When we tolerate misclassification on the border up to 25 data frames, we obtain 98.68\% accuracy, 91.52\% recall, 92.5\% precision, and 91.34\% F1 on average over all activities. The detailed results for each activity are shown in Table \ref{tb:results_bordertol}. 
The confusion matrix obtained with border tolerance is presented in Table \ref{tb:confmatrix_bordertol}.

\begin{table*}[h]
	\centering
	\caption{Continuous activity recognition allowing border tolerance}
	\label{tb:results_bordertol}
	\begin{adjustbox}{max width=\textwidth}
		\begin{tabular}{c|ccccccccc}
			~& Walking & Running & Going up & Going down & Sitting & Sitting down & Standing up & Standing & Average\\  
			\hline
			Recall  	 &94.59 &  97.49 &  94.77  &   91.58   &   98.9 &    74.18    &    85.25   &  96.27  &        91.52	\\	 
			Precision 	 &96.84 &  93.21 &  87.03  &   92.89   &  92.29 &    90.86    &    88.66   &  98.17  &         92.5		\\
			F1 score 	 &95.61 &  94.88 &  90.17  &   91.86   &  95.38 &    80.47    &    85.69   &  97.12  &        91.34 	\\	
			Accuracy    &97.66 &   99.5 &  97.54  &   98.29   &  99.33 &    99.24    &    99.44   &  98.48  &        98.68  \\		
		\end{tabular}
	\end{adjustbox}
\end{table*}

\begin{table*}[h]
	\centering
	\caption{Confusion matrix allowing border tolerance}
	\label{tb:confmatrix_bordertol}
	\begin{adjustbox}{max width=\textwidth}
		\begin{tabular}{c|cccccccc}		
			~& Walking & Running & Going up & Going down & Sitting & Sitting down & Standing up & Standing \\
			\hline
			Walking   &		901806 &  3895  & 22046  &  14510   &   0    &     0      &     41    &   2027   \\
			Running  & 		 1343  & 188584 &  1131  &   1133   &   0    &     0      &     0     &   175   \\
			Going up &		10318  &  1677  & 373222 &   9244   &   0    &     0      &     0     &   351   \\
			Going down & 	 8994  &  5950  & 10887  &  323467  &   0    &     0      &     2     &   611   \\
			Sitting  	&	  0    &   0    &   0    &   387    & 222279 &    519     &    1923   &    0    \\
			Sitting down &	  0    &   0    &   0    &    0     & 11400  &   53175    &    2156   &   5605  \\
			Standing up  &	  0    &   0    &   0    &    0     &  5887  &    675     &   57421   &   3136  \\
			Standing  	&	 7283  &   0    & 20499  &   2994   &   0    &    4671    &    3454   & 1129359 \\
		\end{tabular}	
	\end{adjustbox}	
\end{table*}

In the rest of our experiments, we allowed a border tolerance up to 25 data frames, unless otherwise specified. 

\subsection{Directional features}

\begin{figure}[!ht] 
	\centering
	\includegraphics[width=0.45\textwidth, height=6cm]{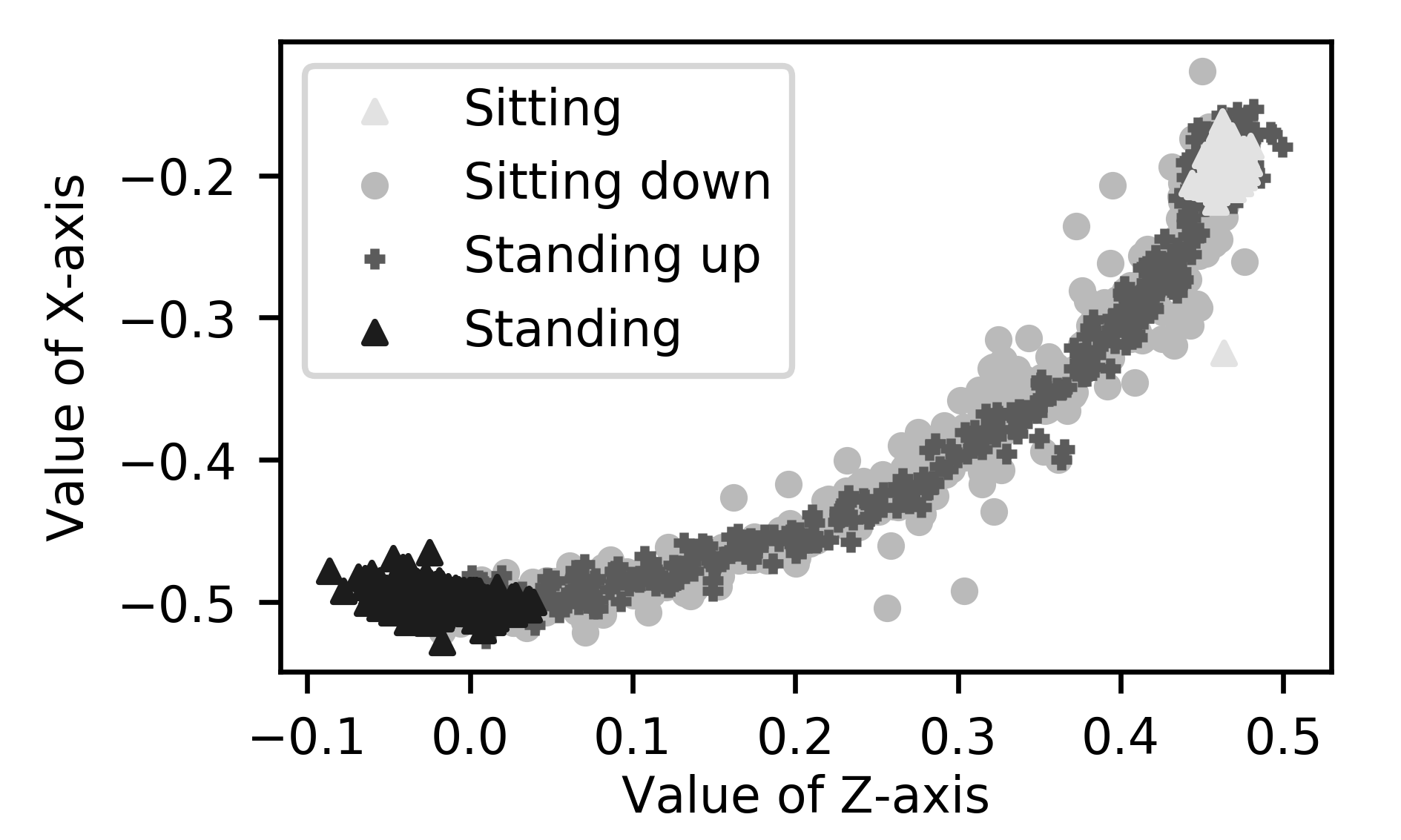}
	\caption{Data from x- and z-axis accelerometer located on left thigh. Data from y-axis accelerometer were nearly constant and thus are not shown. 	\label{fg:sittingacts_3D}}
\end{figure}

Examining the results in Tables \ref{tb:results_bordertol} and \ref{tb:confmatrix_bordertol} shows that the recognition performance of sitting down and standing up activities are relatively poor compared to other activities. We further investigated the problem, and we plotted the data recorded with a 3D accelerometer sensor located on the left thigh during standing, sitting, standing up, and sitting down activities. Data are shown in Figure \ref{fg:sittingacts_3D}. The figure reveals that data from static activities are precisely concentrated on countersides, but the data from dynamic activities lay on top of each other and in-between the static activities. Therefore, it is difficult to distinguish the two dynamic activities. However, if we consider the time stamp of the data in the dynamic activities, we can see that data from the sitting-down activity go from standing to sitting, but data related to the standing-up activity go from sitting to standing activity.   Therefore, we created additional features to indicate changes in signal data. For a signal datum $s_i[t]$ at time $t$ from  $x$ and $z$-axis accelerometer sensors located on both thighs,  we created four additional features as $d_i[t] = s_i[t] - s_i[t - a]$, called directional features, where $i=\{1,2,3,4\}$ indexes the aforementioned signals, and $a$ is a lag parameter denoting time offset. For instance, if $s_i[t]$ is the signal obtained at time $t$ from the $x$-axis accelerometer sensor located on the left thigh, then $d_i[t]$ indicates how much this signal has changed since time $t-a$. Thus, we obtained four additional features.  The original 38-feature-data vector $s[t]$ were concatenated with 4-feature-data vector $d[t]$, yielding 42 features in total for every sample.
These new features add extra information about the direction of movements.

To calibrate the lag parameter, we ran a line search and obtained the best results using $a=15$, which is equivalent to approximately a third of a second (data not shown). Thus, in the rest of our test, we used $a=15$ for the lag parameter. 

The results obtained using the directional features are shown in Table \ref{tb:conf_with_dir_features}, and they indicate a 50--65\% decline in the overall error (cf. Table \ref{tb:results_bordertol}) for the measured metrics. However, closer investigation of the sitting down and standing-up activities reveals even greater improvement. For instance, the F1 score increases from 80.47\% to 93.43\% for sitting down and from 85.69\% to 96.94\% for standing-up. The confusion matrix obtained using directional features, shown in Table \ref{tb:confmatrix_dir_features}, also shows decreased misclassification of activities (cf. Table \ref{tb:confmatrix_bordertol}).

\begin{table*}[h]
	\centering
	\caption{Results of activity recognition with directional features}
	\label{tb:conf_with_dir_features}
	\begin{adjustbox}{max width=\textwidth}
		\begin{tabular}{c|ccccccccc}
			~& Walking & Running & Going up & Going down & Sitting & Sitting down & Standing up & Standing & Average\\
			\hline
			Recall &  96.1  & 98.31  & 95.06   &  91.41    & 99.52  &   90.54     &   96.62    & 98.32   &       95.71   \\  
			Precisions &  97.03  & 93.43  & 91.66   &  94.44    & 96.54  &   96.77     &    97.4    & 99.32   &       95.88  \\  
			F1 scores &  96.46  &  95.4  & 92.74   &  92.46    &  98.0  &   93.43     &   96.94    & 98.81   &       95.55   \\
			Accuracy &  98.13  & 99.58  & 98.33   &  98.49    & 99.72  &   99.71     &   99.86    & 99.41   &       99.15    \\
		\end{tabular}
	\end{adjustbox}
\end{table*}

\begin{table*}[h]
	\centering
	\caption{Confusion matrix using directional features}
	\label{tb:confmatrix_dir_features}
	\begin{adjustbox}{max width=\textwidth}
		\begin{tabular}{c|cccccccc}		
			& Walking & Running & Going up & Going down & Sitting & Sitting down & Standing up & Standing \\
			\hline
			Walking   &		914985 &  2492  & 15590  &   9645   &   0    &     10     &     0     &   1603   \\
			Running  & 		 1028  & 189751 &  580   &   889    &   0    &     0      &     0     &   118   \\
			Going up &		 9280  &  2431  & 374939 &   7932   &   0    &     0      &     0     &   230   \\
			Going down & 	11655  &  6195  &  9536  &  322176  &   0    &     7      &     0     &   342   \\
			Sitting  	&	  0    &   0    &   0    &    0     & 224027 &     26     &    1055   &    0    \\
			Sitting down &	  0    &   0    &   0    &    0     &  7071  &   64733    &    216    &   316   \\
			Standing up  &	  0    &   0    &   0    &    0     &  885   &     0      &   64756   &   1478  \\
			Standing  	&	 5045  &   0    &  3698  &   2554   &   0    &    1973    &    468    & 1154522 \\
		\end{tabular}
	\end{adjustbox}			
\end{table*}

\subsection{State-of-the-art methods}

Here, we introduce the state-of-the-art methods that we used in our comparative tests, and we provide the experimental results of the grid search used to find the best hyperparameter settings. The following methods were used: hidden Markov model (HMM), artificial neural network (ANN), and recurrent neural network (RNN).  HMM was taken from hmmlearn (version 0.2.0), while ANN and RNN were taken from the Keras (version 1.2.2) libraries with Theano (version 0.8.2) support in Python. 

In the HMM, the data emission probabilities were modeled with Gaussian mixture models.  Initial state probabilities were equally 0.125. The state transition probability matrix we used is shown in Table \ref{tb:transmatrix}. Between certain activities, the transition probabilities are set to zero to prohibit absurd transitions. For instance, a {\em sitting} cannot be followed by {\em running} without first {\em standing up}.  We calibrated the transition matrix manually because we did not want HMM to prefer states based on {\em a priori} information obtained from the training data.
 
We ran a grid search on the number of GMM components vs. the window length used in the Viterbi algorithm in order to find the best hyperparameters. Parameters were initialized randomly, and tests were repeated five times. The averaged results (along with the standard deviations (STD) in parentheses) are shown in Table \ref{tb:HMMresults}. Our results indicate that the best accuracy can be achieved using 30 Gaussian components with 50 data frames passed to the Viterbi algorithm. In our experiments with HMMs, we decided to use the same number of GMM components as for the RapidHARe for two reasons: First, this gives us better performance with HMM, and second, the prediction speeds of HMM and RapidHARe becomes comparable. The choice of the window length is also critical. Long windows result in large lag times in prediction. Because the sampling rate is around 56 samples per seconds, the main drawback of long window length is that the system has to wait a long time to collect the adequate number of data samples before prediction. For instance, a window length 50 results in almost a 1 s lag time before any prediction can be made. However, the advantage of long windows is that the prediction can be made for a bigger data chunk, which reduces the prediction time per sample. In our experiments, we decided set the window length to 10 because we found this to be the best trade-off between accuracy and speed. Fewer data yielded worse accuracy, while longer blocks increased the prediction latency.

\begin{table*}
	\centering
	\caption{Transition matrix for hidden Markov model}
	\label{tb:transmatrix}
	\begin{adjustbox}{max width=\textwidth}
		\begin{tabular}{c|cccccccc}
			~& Walking & Running & Going up & Going down & Sitting & Sitting down & Standing up & Standing \\
			\hline
			Walking   & 0.99   &0.0025 &0.0025 &0.0025 &0      &0      &0      &0.0025 \\
			Running   & 0.0025 &0.99   &0.0025 &0.0025 &0      &0      &0      &0.0025 \\
			Going up 	&	0.005  &0      &0.99   &0      &0       &0      &0      &0.005 \\
			Going down  &	0.005  &0      &0      &0.99   &0       &0      &0      &0.005 \\
			Sitting   & 0      &0      &0      &0      &0.99   &0      &0.01  &0      \\
			Sitting down & 0      &0      &0      &0      &0.01   &0.99   &0      &0      \\
			Standing up  & 0      &0      &0      &0      &0      &0      &0.99   &0.01 \\
			Standing  & 0.002  &0.002 &0.002 &0.002 &0      &0.002 &0      &0.99   \\
		\end{tabular}
	\end{adjustbox}	
\end{table*}

\begin{table*}[ht]
	\centering
	\caption{HMM grid search result}
	\label{tb:HMMresults}
	
	{
		\begin{tabular}{cc|cc|cc|c|c|c}
			\multirow{2}{*}{\#GMM} 
			& \multirow{2}{*}{\parbox{1.3cm}{\centering Window length}}		
			& \multicolumn{2}{c|}{Without Border Tolerance } 
			& \multicolumn{2}{c|}{With Border Tolerance} 
			& \multirow{2}{*}{\#Params$^1$}
			& \multirow{2}{*}{\centering Time($\mu$s)$^2$} 
       		& \multirow{2}{*}{\centering Lag(s)$^2$} \\			
			\cline{3-6}
			
			& 
			& \parbox[s]{0.8cm}{\centering F1} 
			& \parbox[s]{1.1cm}{\centering Accuracy} 
			
			& \parbox[s]{0.8cm}{\centering F1}
			& \parbox[s]{1.1cm}{\centering Accuracy} & & \\
			
			\hline		
		
		\hline
		\multirow{4}{*}{30} & 50 & \bf{77.82 (1.02)} & \bf{96.37 (0.30)} & \bf{80.57 (1.08)} & \bf{96.88 (0.31)} & \multirow{4}{*}{18240} & 34.87 (0.04) & 0.85\\
		& 25 & 77.56 (1.00) & 96.32 (0.30) & 80.43 (1.05) & 96.85 (0.31) &  & 54.10 (0.71) & 0.43\\
		& 10 & 76.64 (0.95) & 96.17 (0.30) & 79.36 (0.98) & 96.68 (0.30) &  & 114.05 (0.23) & 0.17\\
		& 5 & 75.78 (0.96) & 95.99 (0.30) & 78.36 (1.00) & 96.47 (0.30) &  & 213.46 (0.28) & 0.09\\
		\hline
		\multirow{4}{*}{20} & 50 & 77.33 (0.38) & 96.03 (0.04) & 80.04 (0.34) & 96.53 (0.05) & \multirow{4}{*}{12160} & 31.39 (0.40) & 0.85\\
		& 25 & 76.93 (0.39) & 95.96 (0.04) & 79.75 (0.36) & 96.47 (0.05) &  & 47.70 (0.04) & 0.43\\
		& 10 & 76.15 (0.32) & 95.80 (0.03) & 78.86 (0.28) & 96.29 (0.04) &  & 105.59 (0.21) & 0.17\\
		& 5 & 75.19 (0.27) & 95.57 (0.03) & 77.73 (0.25) & 96.03 (0.05) &  & 195.65 (0.06) & 0.09\\
		\hline
	
		\multirow{4}{*}{10} & 50 & 76.87 (0.74) & 95.78 (0.10) & 79.58 (0.71) & 96.22 (0.09) & \multirow{4}{*}{6080} & 23.92 (0.05) & 0.85\\
		& 25 & 76.44 (0.77) & 95.73 (0.11) & 79.22 (0.74) & 96.16 (0.09) &  & 42.01 (0.03) & 0.43\\
		& 10 & 75.57 (0.72) & 95.59 (0.12) & 78.27 (0.66) & 96.01 (0.10) &  & 93.41 (0.21) & 0.17\\
		& 5 & 74.81 (0.73) & 95.41 (0.11) & 77.36 (0.68) & 95.81 (0.10) &  & 179.37 (0.06) & 0.09\\
		\hline
		\multirow{4}{*}{5} & 50 & 75.28 (0.23) & 94.87 (0.11) & 77.92 (0.24) & 95.29 (0.11) & \multirow{4}{*}{3040} & 21.42 (0.11) & 0.85\\
		& 25 & 74.68 (0.20) & 94.79 (0.09) & 77.21 (0.21) & 95.20 (0.10) &  & 39.04 (0.05) & 0.43\\
		& 10 & 73.28 (0.15) & 94.57 (0.07) & 75.64 (0.16) & 94.94 (0.08) &  & 88.38 (0.01) & 0.17\\
		& 5 & 72.31 (0.12) & 94.30 (0.01) & 74.52 (0.11) & 94.63 (0.00) &  & 171.12 (0.50) & 0.09\\
		\hline
		
	\end{tabular}
}\\
\raggedright
Tests were repeated five times; mean results are shown along with STD in parentheses. Performance measures are averaged over activities. $^1$The number of parameters in the models to be trained. $^2$Time in micro seconds to predict the activity of a single data frame measured on a single-thread CPU. $^3$ Time in seconds to wait to collect an  adequate number of data samples.

\end{table*}

\begin{figure}[tbp] 
	\centering
	\includegraphics[width=7cm,height=5cm]{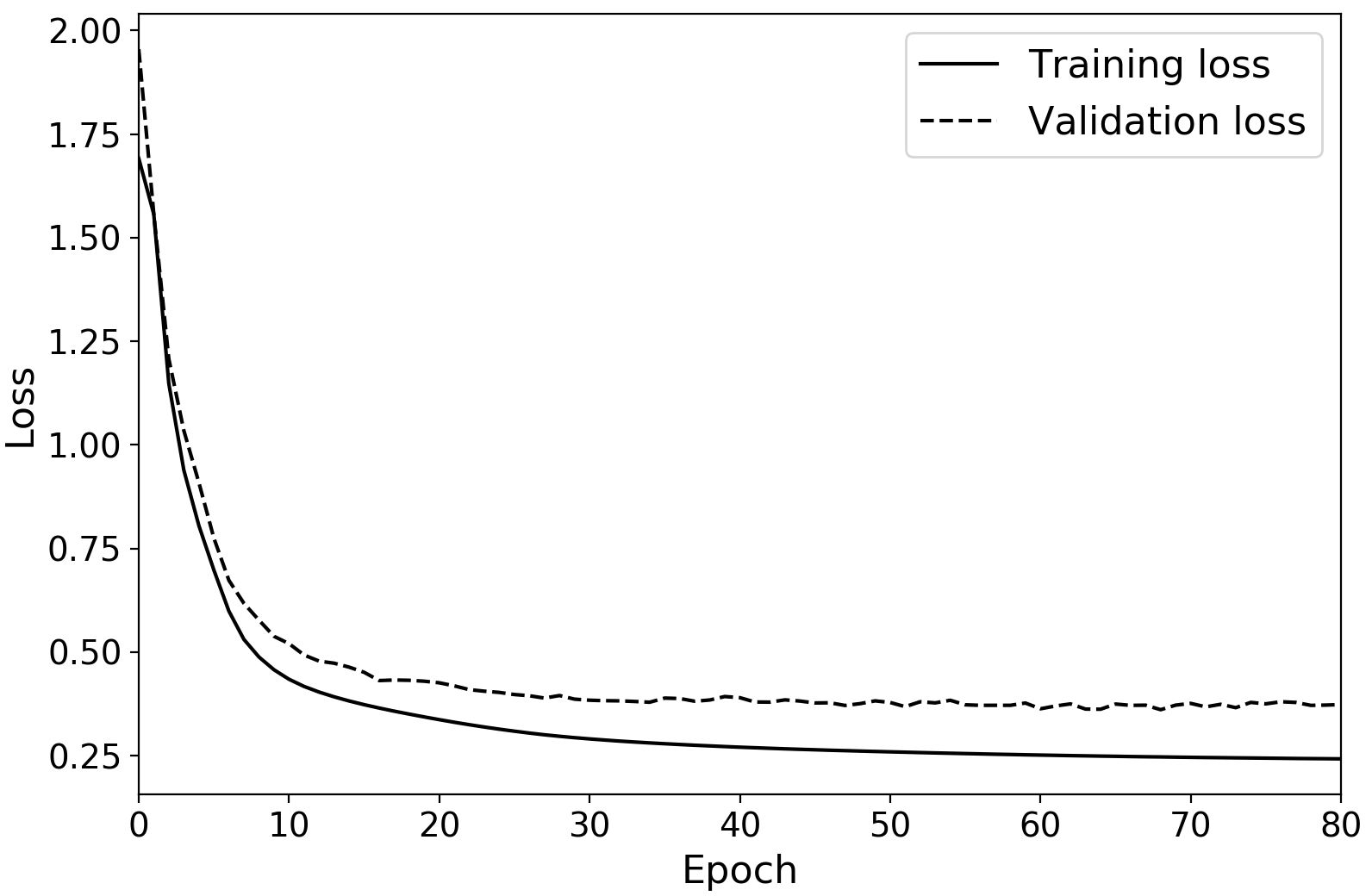}
	\caption{Learning curve for early stopping. Training terminated after epoch 80 because of lack of improvement on the validation set.}
	\label{fg:learning-curve}
\end{figure}

To find the best ANN structure, we ran a grid search over the following hyperparameters: (1) number of hidden units within a layer from 10 to 400; (2) number of hidden layers: 1 or 2; (3) activation function: sigmoid or rectified linear unit (ReLU). The training was performed with an Adam optimizer and with early stopping. In the early stopping, the training stopped if the validation loss reduced less than 1e-6 in the last three epochs or if the average validation loss of the last 10 epochs was greater than the average validation loss of the preceding 10 epochs (that is, the cost tended to grow). An example for the learning curves along with the loss on the validation set is shown in Figure \ref{fg:learning-curve}. Tests were repeated five times; average results are shown in Table \ref{tb:ANNresults}, along with STD in parentheses. The results indicate that structures with the ReLU activation function performed poorly; however, two--layered structure with a sigmoid activation function seemed to be overfit and slow in prediction. The best performance with ANN can be achieved using a single layer network with sigmoid activation function having 200 hidden unites, and this is the structure we used in our comparative tests.

\begin{table*}[ht]
	\centering
	\caption{Artificial neural network grid search result}
	\label{tb:ANNresults}
	
	{
		\begin{tabular}{c|cc|cc|c|c}
			\multirow{2}{*}{\#Units} 
			& \multicolumn{2}{c|}{Without Border Tolerance } 
			& \multicolumn{2}{c|}{With Border Tolerance} 
			& \multirow{2}{*}{\#Params$^1$}
			& \multirow{2}{*}{\parbox{1.3cm}{\centering Time ($\mu$s)$^2$}} \\
			
			\cline{2-5}			
			
			& \parbox[s]{0.8cm}{\centering F1} 
			& \parbox[s]{1.1cm}{\centering Accuracy} 
			
			& \parbox[s]{0.8cm}{\centering F1}
			& \parbox[s]{1.1cm}{\centering Accuracy} & & \\
			
			\hline		
		
		\hline
		\multicolumn{6}{c}{Two hidden layers with sigmoid activation function} \\
		\hline
		400 & 85.34 (0.27) & 97.78 (0.04) & 87.65 (0.29) & 98.1 (0.04)  & 179208  & 45.27 (0.77) \\
		300 & 85.86 (0.46) & 97.82 (0.02) & 88.23 (0.56) & 98.13 (0.04) & 104408  & 34.51 (0.4) \\
		200 & 85.45 (0.41) & 97.8 (0.04) & 87.77 (0.34) & 98.11 (0.03)  & 49608   & 24.03 (0.11) \\
		100 & 86.4 (0.09) & 97.84 (0.01) & 88.82 (0.14) & 98.16 (0.02)  & 14808   & 13.51 (0.24) \\
		50 & 86.5 (0.63) & 97.77 (0.08) & 88.93 (0.68) & 98.08 (0.08)   & 4908	   & 8.35 (0.31) \\
		20 & 85.22 (0.23) & 97.4 (0.03) & 87.59 (0.25) & 97.67 (0.03)   & 1368	   & 4.74 (0.07) \\
		10 & 80.15 (0.35) & 96.44 (0.06) & 81.97 (0.37) & 96.63 (0.06)  & 588     & 3.69 (0.05) \\

		\hline
		\multicolumn{6}{c}{One hidden layer with sigmoid activation function} \\
		\hline
		400 & 86.73 (0.39) & 97.88 (0.04) & 89.09 (0.49) & 98.18 (0.04) & 18808& 23.2 (0.16) \\
		300 & 86.97 (0.23) & 97.88 (0.03) & 89.3 (0.27) & 98.18 (0.04)  & 14108& 18.52 (0.43) \\
		200 & \bf{86.98 (0.04)} & \bf{97.88 (0.02)} & \bf{89.4 (0.08)} & \bf{98.19 (0.01)}  & 9408&  13.01 (0.04) \\
		100 & 86.51 (0.19) & 97.78 (0.03) & 88.79 (0.26) & 98.07 (0.04) & 4708&  7.77 (0.08) \\
		50 & 85.77 (0.28) & 97.62 (0.04) & 88.1 (0.3) & 97.89 (0.04)    & 2358&  4.9 (0.05) \\
		20 & 83.19 (0.55) & 97.07 (0.05) & 85.3 (0.68) & 97.31 (0.06)   & 948 &  3.48 (0.05) \\
		10 & 77.69 (0.23) & 95.89 (0.04) & 79.37 (0.2) & 96.05 (0.04)   & 478 &  3.02 (0.14) \\
		\hline
		\multicolumn{6}{c}{Two hidden layers with ReLU activation function} \\
		\hline
		400 & 82.37 (0.19) & 97.35 (0.03) & 84.51 (0.15) & 97.66 (0.03) & 179208  & 8.09 (0.06) \\
		300 & 81.75 (0.15) & 97.23 (0.03) & 83.95 (0.15) & 97.54 (0.02) & 104408  & 6.63 (0.04) \\
		200 & 81.73 (0.22) & 97.22 (0.08) & 83.91 (0.24) & 97.54 (0.08) & 49608   & 5.33 (0.02) \\
		100 & 81.66 (0.87) & 97.23 (0.11) & 83.92 (0.88) & 97.55 (0.11) & 14808   & 4.33 (0.16) \\
		50 & 81.74 (0.24) & 97.23 (0.01) & 84.04 (0.27) & 97.54 (0.01) &  4908	 & 3.77 (0.03) \\
		20 & 82.7 (0.75) & 97.16 (0.09) & 84.99 (0.84) & 97.44 (0.09)  &  1368	 & 3.23 (0.04) \\
		10 & 78.27 (0.62) & 96.25 (0.09) & 80.15 (0.63) & 96.44 (0.09) &  588    & 3.13 (0.01) \\
		\hline
		\multicolumn{6}{c}{One hidden layer with ReLU activation function} \\
		\hline
		400 & 83.29 (0.48) & 97.45 (0.04) & 85.56 (0.5) & 97.77 (0.03)  & 18808  & 5.57 (1.32) \\
		300 & 83.2 (0.33) & 97.43 (0.02) & 85.38 (0.38) & 97.74 (0.03)  & 14108  & 4.14 (0.02) \\
		200 & 83.7 (0.42) & 97.49 (0.03) & 85.92 (0.43) & 97.79 (0.03)  & 9408   & 3.78 (0.1) \\
		100 & 85.07 (0.26) & 97.56 (0.06) & 87.34 (0.25) & 97.85 (0.06) & 4708   & 3.38 (0.06) \\
		50 & 84.9 (0.41) & 97.46 (0.05) & 87.16 (0.44) & 97.73 (0.05)   & 2358   & 2.9 (0.08) \\
		20 & 82.67 (0.37) & 97.0 (0.03) & 84.79 (0.34) & 97.24 (0.03)   & 948	  & 2.78 (0.04) \\
		10 & 77.97 (0.49) & 96.13 (0.08) & 79.63 (0.5) & 96.29 (0.08)   & 478    & 2.76 (0.04) \\
	\end{tabular}
}
\end{table*}

For the best hyperparameter search for the RNN, we ran a grid search over the number of  hidden units from 10 to 200 using sigmoid or ReLU activation functions. Tests were repeated five times, and the averaged results along with STD presented in Table \ref{tb:RNNresults1000}. The results indicate that RNN can be considered rather slow. Moreover, ReLU seems to perform poorly compared to the sigmoid activation function. The best performance was achieved using 200 hidden units with a sigmoid activation function organized in a single layer. Thus, this is the structure for RNN we used in our comparative tests. 

\begin{table*}[ht]
	\centering
	\caption{Recurrent Neural network}
	\label{tb:RNNresults1000}
		\begin{tabular}{c|cc|cc|c|c}
			\multirow{2}{*}{\#Units} 
			& \multicolumn{2}{c|}{Without Border Tolerance } 
			& \multicolumn{2}{c|}{With Border Tolerance} 
			& \multirow{2}{*}{\#Params$^1$}
			& \multirow{2}{*}{\parbox{1.3cm}{\centering Time ($\mu$s)$^2$}} \\
			
			\cline{2-5}

			& \parbox[s]{0.8cm}{\centering F1} 
			& \parbox[s]{1.1cm}{\centering Accuracy} 
			
			& \parbox[s]{0.8cm}{\centering F1}
			& \parbox[s]{1.1cm}{\centering Accuracy} & & \\
			
			\hline		
						
			\hline

		\multicolumn{6}{c}{One layer with sigmoid activation function} \\
		\hline
		200 & \bf{82.97 (1.28)} & \bf{97.58 (0.16)} & \bf{83.31 (1.29)} & \bf{97.64 (0.16)} & 17208  & 149.854 (1.189) \\
		150 & 80.15 (1.89) & 97.21 (0.36) & 80.57 (1.85) & 97.29 (0.36) & 12908  & 123.09 (1.83) \\
		100 & 78.93 (1.42) & 96.79 (0.46) & 79.29 (1.47) & 96.86 (0.46) & 8608   & 94.25 (0.44) \\
		75 & 82.09 (2.09) & 97.44 (0.23) & 82.43 (2.09) & 97.52 (0.23)  & 6458   & 82.33 (1.25) \\
		50 & 75.54 (1.43) & 96.57 (0.33) & 75.88 (1.47) & 96.64 (0.34)  & 4308   & 69.35 (1.36) \\
		20 & 73.83 (2.86) & 96.49 (0.3) & 74.12 (2.86) & 96.55 (0.31)   & 1728   & 52.25 (0.07) \\
		10 & 70.81 (2.96) & 95.73 (0.56) & 71.13 (2.99) & 95.79 (0.57)  & 868   & 43.05 (1.15) \\

		\hline
		\multicolumn{6}{c}{One layer with ReLU activation function} \\
		\hline
		200 & 65.0 (6.15) & 92.39 (1.59) & 65.27 (6.2) & 92.45 (1.59)   & 17208  & 61.54 (0.36) \\
		150 & 74.02 (5.01) & 95.03 (1.36) & 74.35 (5.08) & 95.09 (1.37) & 12908  & 56.8 (0.73) \\
		100 & 72.06 (3.05) & 94.33 (0.79) & 72.35 (3.07) & 94.39 (0.79) & 8608 & 53.05 (1.0) \\
		75 & 69.34 (3.48) & 94.02 (0.79) & 69.58 (3.53) & 94.08 (0.8)   & 6458 & 52.25 (1.21) \\
		50 & 70.19 (1.53) & 93.94 (0.54) & 70.44 (1.53) & 94.0 (0.54)   & 4308 & 49.86 (0.55) \\
		20 & 61.68 (6.27) & 93.64 (1.12) & 61.93 (6.31) & 93.7 (1.12)   & 1728 & 46.73 (1.49) \\
		10 & 34.13 (4.54) & 87.86 (0.97) & 34.25 (4.56) & 87.9 (0.98)   & 868& 42.58 (0.87) \\
	\end{tabular}
\end{table*}

\subsection{Comparison to state-of-the-art methods}

\begin{table*}[ht]
	\centering
	\caption{A) Main classification results}
	\label{tb:results}
	
	{\setlength{\tabcolsep}{5pt}
		\begin{tabular}{c|cccc|cccc|c|c}
			\multirow{2}{*}{Method} 
			& \multicolumn{4}{c|}{Without Border Tolerance } 
			& \multicolumn{4}{c|}{With Border Tolerance} 
			& \multirow{2}{*}{\#Params$^2$}
			& \multirow{2}{*}{\parbox{1.3cm}{\centering Time ($\mu$s)$^3$\\(std)}} \\
			
			\cline{2-9}
			
			& \parbox[s]{0.8cm}{\centering Recall\\(std)}
			& \parbox[s]{1.0cm}{\centering Precision\\(std)}
			& \parbox[s]{0.8cm}{\centering F1\\(std)} 
			& \parbox[s]{1.1cm}{\centering Accuracy\\(std)} 
			
			& \parbox[s]{0.7cm}{\centering Recall\\(std)}
			& \parbox[s]{1.0cm}{\centering Precision\\(std)}
			& \parbox[s]{0.8cm}{\centering F1\\(std)}
			& \parbox[s]{1.1cm}{\centering Accuracy\\(std)} & & \\
			
			\hline 
			\parbox[s]{1.5cm}{\multirow{2}{*}{\centering RapidHARe}}  
			& 87.14 & 87.75 & 86.48 & 97.83 & 91.44 & 92.82 & 91.4  & 98.65& \multirow{2}{*}{6536} & \bf{9.7}\\
			& (0.23)  & (0.14)  & (0.2)   & (0.05)  & (0.31)  & (0.17)  & (0.27)  & (0.05) &  & (0.032)\\

			\multirow{2}{*}{\parbox[s]{2cm}{\centering RapidHARe-DF$^1$}}
			
			& \bf{88.93} & 88.01 & \bf{87.9} & \bf{97.92} & \bf{94.59} & \bf{94.58} & \bf{94.27} &\bf{ 98.94} &\multirow{2}{*}{9064} &  9.933\\ 
			& 0.12  & 0.13  & 0.14 & 0.02  & 0.13  & 0.12  & 0.14  & 0.03  & &  0.219\\

			\parbox[s]{1cm}{\multirow{2}{*}{\centering ANN}}             

			& 84.55 & \bf{89.01} & 86.98 & 97.88 & 86.82 & 91.13 & 89.4 & 98.19& \multirow{2}{*}{9408}  & 13.01\\
			& 0.03  & 0.06  & 0.04  & 0.02  & 0.01  & 0.09  & 0.08  & 0.01 &  & 0.04\\
			
			\parbox[s]{1cm}{\multirow{2}{*}{\centering RNN} }

			 & 80.87 & 87.21 & 82.97 &  97.58  & 81.11 & 87.59 & 83.31 & 97.64& \multirow{2}{*}{17208}  & 149.854\\
			 & 1.33  & 0.36  & 1.28   & 0.16   & 1.34  & 0.34  & 1.29   & 0.16 &  & 1.189\\
						                                 
			\parbox[s]{1cm}{\multirow{2}{*}{\centering HMM}}

			 & 82.59 & 83.14 & 81.54 & 96.76 & 85.03 & 86.11 & 84.34 & 97.24& \multirow{2}{*}{6536} & 102.862\\
			& 0.11  & 0.09  & 0.1   & 0.02  & 0.14  & 0.08  & 0.12  & 0.01 &  & 0.184\\
			\hline

\multicolumn{11}{c}{B) Results using only triaxial accelerometer data obtained from thigh and shin}\\

			\hline 
			
			\multirow{2}{*}{\parbox[s]{1.5cm}{\centering RapidHARe}} 
     		& 75.84 & 74.04 & 72.32 & 96.21 & 79.6  & 78.11 & 75.89 & 96.97& \multirow{2}{*}{2064} & \bf{8.859}\\
			& 0.13  & 0.09  & 0.13  & 0.09  & 0.07  & 0.1   & 0.09  & 0.08 &  & 0.091\\
			
			\multirow{2}{*}{\parbox[s]{2cm}{\centering RapidHARe-DF$^1$}}   
			& \bf{89.33} & \bf{87.19} & \bf{87.64} & \bf{97.93} & \bf{94.43} & \bf{94.21} & \bf{94.04} & \bf{98.92} & \multirow{2}{*}{3136} & 9.183\\
			& 0.02  & 0.14  & 0.09  & 0.0   & 0.02  & 0.11  & 0.06  & 0.0  &  & 0.17\\
			
			\multirow{2}{*}{\parbox[s]{1cm}{\centering ANN} }           
			& 70.4  & 74.74 & 70.36 & 95.84 & 70.86 & 75.35 & 70.82 & 95.96& \multirow{2}{*}{4208	} & 12.759\\
			& 0.32  & 0.11  & 0.41  & 0.05  & 0.31  & 0.12  & 0.41  & 0.05 &  & 0.071\\
			
			\multirow{2}{*}{\parbox[s]{1cm}{\centering RNN} }
			& 79.71 & 85.87 & 80.6  & 97.27 & 79.95 & 86.3  & 80.94 & 97.35& \multirow{2}{*}{6808
			}  & 154.799\\
			& 0.94  & 0.55  & 0.91  & 0.18  & 0.96  & 0.59  & 0.94  & 0.18 &  & 6.158\\
			
			\multirow{2}{*}{\parbox[s]{1cm}{\centering HMM} }        

			& 75.23 & 73.23 & 71.61 & 96.0  & 78.03 & 75.95 & 74.13 & 96.53& \multirow{2}{*}{2064} & 95.571\\
			& 0.1   & 0.13  & 0.12  & 0.09  & 0.04  & 0.12  & 0.09  & 0.09 &  & 0.032\\
			\hline
	
\multicolumn{11}{c}{C) Results using only triaxial accelerometer data obtained from thigh}\\
	
			\hline 
			
			\multirow{2}{*}{\parbox[s]{1.5cm}{\centering RapidHARe}}
         	& 76.98 & 76.94 & 74.45 & 96.56 & 79.94 & 80.92 & 77.58 & 97.28& \multirow{2}{*}{1032}  & \bf{8.724}\\
			& 0.02  & 0.02  & 0.03  & 0.0   & 0.01  & 0.03  & 0.02  & 0.0  &  & 0.021\\
			
		\multirow{2}{*}{	\parbox[s]{2cm}{\centering RapidHARe-DF$^1$} }
			& \bf{88.6}  & \bf{87.3}  & \bf{87.33} & \bf{97.89} & \bf{93.21} & \bf{94.19} & \bf{93.49} & \bf{98.9}& \multirow{2}{*}{1768} & 8.945\\
			& 0.0   & 0.02  & 0.02  & 0.0   & 0.0   & 0.02  & 0.01  & 0.0  &  & 0.094\\
			
			\multirow{2}{*}{\parbox[s]{1cm}{\centering ANN}  }         
			& 61.89 & 67.16 & 61.93 & 93.73 & 62.12 & 67.47 & 62.16 & 93.8 & \multirow{2}{*}{3008
			} & 12.188\\
			& 0.5   & 0.3   & 0.51  & 0.04  & 0.48  & 0.28  & 0.51  & 0.04 &  & 0.025\\
			
			\multirow{2}{*}{\parbox[s]{1cm}{\centering RNN} }

			& 75.55 & 81.75 & 76.35 & 96.12 & 75.86 & 82.16 & 76.73 & 96.23& \multirow{2}{*}{4408
			} & 150.192\\
			& 2.85  & 3.51  & 3.31  & 0.52  & 2.87  & 3.53  & 3.35  & 0.54 &  & 1.27\\
			
			\multirow{2}{*}{\parbox[s]{1cm}{\centering HMM} }       

			& 73.8  & 73.67 & 70.96 & 95.63 & 76.35 & 76.88 & 73.54 & 96.29& \multirow{2}{*}{1032} & 93.944\\
			& 0.01  & 0.02  & 0.0   & 0.0   & 0.02  & 0.06  & 0.02  & 0.0  &  & 0.164\\
			\hline
		\end{tabular}
	}
\\
	Performance measures are averaged over activities. $^1$RapidHARe using directional features (DF). $^2$The number of parameters in the models to be trained. $^3$Time in micro seconds to predict the activity of a single data frame measured on a single-thread CPU.
\end{table*}

Here we compare the performance of the  RapidHARe methods to state-of-the-art methods. Recognition performance was evaluated by recall, precision, F1 score, and accuracy, and our main results are summarized in the Table \ref{tb:results}A. The best results were achieved using the RapidHARe method using directional features (RapidHARe-DF) and all features from all sensors when we allowed tolerance on the border between activities. RapidHARe-DF has achieved a 94.27\% F1 score and 98.94\% accuracy. Compared to ANN, RNN, and HMM, this decreased the F1 score error rate by 46\%, 66\%, and 63\% and the accuracy error rate by 41\%, 55\%, and 62\%,  respectively. Allowing border tolerance improves performance metrics. For instance, by allowing border tolerance, the RapidHARe-DF method reduced the F1 score error rate by 52\% and the accuracy error rate by 49\% when compared to the case when border tolerance was not allowed. However, border tolerance for ANN, RNN, and HMM reduced the F1-score error rate by 19\%, 2\%, and 15\%, respectively, and the accuracy error rate by 15\%, 2\%, and  15\%, respectively. This suggests that the ANN, RNN, and HMM methods tend to make more scattered misclassifications within the same activity rather than at the border between different activities.

Because one of our aims is to develop a simple model for HAR prediction, we tested these methods with fewer features as well. First, we kept the triaxial accelerometer data obtained from accelerometers located on the thigh and shin, and second, we kept the accelerometer data from only the thigh. All gyroscope and EMG data were omitted. The results are shown in Tables \ref{tb:results}B and \ref{tb:results}C. When border tolerance is taken into account, ANN's performance  drops from 89.4\% to 62.16\% in the F1 score as the amount of information and the number of  features decrease. 
The F1 scores for RNN and HMM decrease moderately from 83.31\% to 76.73\% and 84.34\% to 73.54\%, respectively. 
While RapidHARe also shows loss in performance, RapidHare-DF seems to be robust, and its performance remains roughly the same; it outperforms all state-of-the-art methods under limited data. Similar tendencies can be observed when the performance is evaluated in accuracy with and without allowing border tolerance. 

The CPU time is remarkably low for our model. RapidHARe and RapidHARe-DF perform activity predictions around one and a half times faster than ANN, eight times faster than HMM , and more than ten times faster than RNN. It is worth noting that the number of model parameters is also the lowest for our model, while HMM and RNN consist of significantly more parameters. The timing results and the number of parameters are shown in Table \ref{tb:results} as well. In our opinion, these facts make our model plainly appropriate for real-time recognition.

\section{Conclusions}
In this article, we have presented a new, fast, and computationally inexpensive method, called RapidHARe, for continuous activity recognition. It predicts activities based on the distribution of the raw data in a small, half-second-long context window, in which the distribution was modeled using Gaussian mixture models. Note that, our method does not employ any dynamic-programming-based algorithms for inference, as they are known to be slow. This fact makes RapidHARe an extremely fast predictor; as comparative tests showed, our method is one and a half times faster than an ANN method, and more than eights time faster than RNN and HMM methods. 

RapidHARe outperforms the current state-of-the-art methods in accuracy as well. However, performance can be further improved using additional features, termed directional features, that exploit information about signal changes. This information is especially useful in distinguishing among sitting-related activities, such as sitting down and standing up. We also discussed the difficulty of exactly determining the border between two subsequent activities in the signal. If we allow a little tolerance around the border in the performance evaluation, then RapidHARe provides nearly perfect performance, while the other methods' performance remain roughly the same. This, in our opinion, indicates that the other methods tend to make scattered misclassifications within the same activity. 

It is also worth mentioning that our method did not utilize any data preprocessing, feature-selection, extraction, or transformation methods, and it still achieved outstanding performance. Perhaps these preprocessing methods could contribute to better performance, but this would come at the expense of additional CPU time. 

In this article, we investigated HAR methods from purely computational aspects, but we did not discuss any hardware-related issues or how our systems could be implemented on mobile devices. Since our method is the fastest and requires the smallest amount of memory to store the predictor model, we believe that RapidHARe would consume the least amount of energy compared to the current state-of-the-art methods, independently from the hardware specifications. That is, if a HAR system were implemented on a PC, mobile phone, or microcontroller, the energy consumption for data collection or for wireless data transfer from the sensors to work stations (PC, mobile phone) would be the same independently from the chosen HAR model. 



Finally, we also mention that GPUs  (and NPUs) are becoming standard chips in mobile devices in order to perform AI features $-$ for instance, in Huawei's Mate 10 (Kirin 970) and Google's Pixel 2 (Adreno 540) $-$ and therefore, HAR systems could perform inference on these GPUs. In this case, the inference will become faster and independent of the method, albeit at the expense of additional energy consumption. As we argued in the introduction that the HAR problem is simple and does not require a large number of data features and computationally exhaustive inference algorithms, we think that the speed gained by GPUs might be not worth the additional energy consumption required by GPUs and by the data transfer from the CPU/memory to the GPU.



	

\section*{Acknowledgements}
	We gratefully acknowledge the support of NVIDIA Corporation with the donation of the GTX Titan X GPU used for model parameter training in this research. We would like thank the participants in the data acquisition for the effort and time they devoted to this work. We would also like to thank to Timur Bergaliyev and his lab members Sergey Sakhno and Sergey Kravchenko  from Laboratory of Applied Cybernetic Systems at MIPT and BiTronics Lab (\url{www.bitronicslab.com}) for their technical support on using sensors. 



\bibliographystyle{plain}           
\bibliography{bibliography}        

\end{document}